\documentclass{pasj01}
\Received{$\langle$reception date$\rangle$}
\Accepted{$\langle$acception date$\rangle$}
\Published{$\langle$publication date$\rangle$}

\usepackage{ulem}

\begin{document}

\title{Property of Young Massive Clusters in a Galaxy-Galaxy Merger Remnant}

\author{Hidenori Matsui${}^{1,2}$, Ataru Tanikawa${}^{2,3}$, and Takayuki R Saitoh${}^4$}%
\altaffiltext{1}{National Institute of Technology, Asahikawa College, Asahikawa, Shunkodai 2-2-1-6, Asahikawa, Hokkaido, 071-8142, Japan}
\altaffiltext{2}{Department of Earth Science and Astronomy, College of Arts and Sciences, The University of Tokyo, 3-8-1 Komaba, Meguro-ku, Tokyo, 153-8902, Japan}
\altaffiltext{3}{RIKEN Advanced Institute for Computational Science, 7-1-26 Minatojima-minami-machi, Chuo-ku, Kobe, Hyogo 650-0047, Japan}
\altaffiltext{4}{Earth-Life Science Institute, Tokyo Institute of Technology, 2-12-1, Ookayama, Meguro, Tokyo, 152-8551, Japan}
\email{matsui@asahikawa-nct.ac.jp}


\maketitle

\begin{abstract}
We investigate the properties of young massive clusters (YMCs) in a galaxy-galaxy merger remnant by analyzing the data obtained by a gas rich major merger simulations in Matsui et al. 2012.
We found that the YMCs are distributed at a few $\rm kpc$ and at $\sim 10~{\rm kpc}$ from the galactic center, in other words, there are two components of their distribution.
The former are formed in filamentary and turbulent gas generated at a few $\rm kpc$ from the center because of galaxy encounters, and the latter are formed in tidal tails which are far from the center.
The YMCs are much less concentrated than galaxy stars.
The mass function of the YMCs is $dN/dM \propto M^{-2}$.
Most of YMCs are formed from the second encounter to the final coalescence phase of the galactic cores,
and their formation rate is especially high at final coalescence phase.
Most of them consists of single stellar population in age, but YMCs with multi stellar populations in age are also formed.
The multiple populations are produced by the following process:
a YMC captures dense gas, and another generation stars form within the cluster.
There are several YMCs formed in an isolated disk before the encounter of galaxies.
These candidates contain stars with various age by capturing dense gas and forming stars.
YMCs in a merger remnant, have various orbits, but large fraction of candidates have circular orbits.
\end{abstract}

\section{Introduction}

Major mergers of gas-rich galaxies are expected to play an important role in formation of globular clusters.
The CO observation of local interacting galaxies showed a compressed gas filament, which may lead to formations of globular clusters, induced by galaxy-galaxy collision \citep{2018arXiv180511630K}.
The optical and infrared images found young massive star clusters with $\sim 10^{5-6}~{M_{\odot}}$ formed in galaxy-galaxy merging process \citep{1995AJ....109..960W,1999AJ....118.1551W,2008A&A...489.1091M}.
In galaxy-galaxy merger remnants, globular cluster systems formed in a galaxy-galaxy merging process are detected \citep{2017MNRAS.466.4259B,2018ApJ...859..108K}.
In order to study formation, evolution, and properties of such globular cluster systems in more detail, numerical simulations of a galaxy-galaxy merger are needed.

Previously, a large number of numerical simulations of the galaxy mergers have been performed \citep{1996ApJ...464..641M,1996ApJ...471..115B,2005ApJ...623L..67K,2006ApJ...650..791C,2007A&A...468...61D}.
These numerical simulations succeeded to understand roughly dynamics in galaxy-galaxy merging process such as gas inflow to the galactic central region or a central starburst.
These studies, however, were difficult to reproduce shock-induced star and star cluster formations in the merging process as shown by observations.
This is because they used unrealistic interstellar medium (ISM) model with temperature cutoff at $T= 10^4 ~\rm{K}$ in a cooling function due to the limited mass and spatial resolution.
The artificial high temperature floor of ISM prevents ISM from gravitational instability at the shock generated by the galaxy-galaxy encounter.

In order to investigate formations of star clusters in galaxy-galaxy merging process, subgrid model for star clusters is used instead of resolving formations of individual star clusters directly \citep{2002MNRAS.335.1176B,2012MNRAS.421.1927K}.
\citet{2008MNRAS.389L...8B} have performed high-resolution simulations of merging galaxies and reproduced formations of star clusters directly, although they used method of sticky particles instead of resolving hydrodynamics.

Recently, the higher mass and spatial resolution simulations of merging galaxies have been performed \citep{2009PASJ...61..481S,2009ApJ...694L.123K,2010ApJ...720L.149T,2012ApJ...746...26M,2013MNRAS.430.1901H,2013MNRAS.433...78H,2015MNRAS.446.2038R}.
The high resolution simulations allow us to take radiative cooling of low temperature gas ($T<10^4~{\rm K}$) into account and realize multi-phase nature of ISM.
These simulations have naturally succeeded to reproduce shock-induced star and star cluster formations.
These studies have clarified shock-induced star cluster formations at the first encounter of galaxies \citep{2009PASJ...61..481S,2010ASPC..423..185S},
formations of hypermassive star clusters in galaxy-galaxy merging process \citep{2012ApJ...746...26M},
and formation mechanism of star cluster at each interaction stage \citep{2015MNRAS.446.2038R}.
The formation of globular clusters in merging galaxies at high-redshift are also clarified by cosmological simulations \citep{2018MNRAS.474.4232K}.
Although formations of star clusters in the merging process were clarified, properties of globular clusters in a merger remnant have not been analyzed yet.

In this paper, we focus on and investigate properties of young massive cluster (YMCs) formed in galaxy-galaxy merging process, which can produce luminous infrared galaxies (LIRGs) as shown in paper I.
For this purpose, we analyze a galaxy-galaxy merger remnant obtained by \citet{2012ApJ...746...26M} (hereafter paper I).
We describe our methods in \S 2 and results in \S 3.
Summary and discussions are presented in \S 4.

\section{Method}

\subsection{Simulation data}

To investigate a merger remnant and YMCs in the remnant, simulation data obtained by paper I is analyzed.
Here, the simulations have been performed by Tree+GRAPE SPH/N-body code ``ASURA'' \citep{2008PASJ...60..667S}.
We analyze mainly the data of the highest resolution model called $H_{\rm TT,5pc}$.
In this model, initial setup is as follows.
Firstly, we simulate an isolated disk galaxy, which consists of an exponential disk with $6.3\times 10^{9}~M_{\odot}$ and a dark matter halo with $1.1\times 10^{11}~M_{\odot}$, for $1000~{\rm Myr}$.
The initial disk and halo include gas, of which metallicity is $0.01$, with $1.2\times 10^{9}~M_{\odot}$ and $1.1\times 10^{9}~M_{\odot}$, respectively.
The initial metallicity is slightly less than solar metallicity and larger than that of large magellanic cloud \citep{2016MNRAS.455.1855C}.
After stabilizing the disk, we start galaxy-galaxy merger simulations in which both disks are tilted at $-109^{\circ}$ and $71^{\circ}$ to the orbital plane, respectively, and prograde-prograde encounter occurs.
We set the time of starting the merger simulation to be $t=0$.
At $t=0$, each galaxy includes gas with $1.5\times 10^9~M_{\odot}$ since gas decreases due to star formation.
Hereafter, initial stars at $t=-1000~{\rm Myr}$, newly born stars before $t=0$, and after $t=0$ are called ``pre-existing stars'', ``old stars'', and ``new stars''.
After starting the merger simulation, the first, second, and third encounters occur at $t\sim 450~{\rm Myr}$, $t\sim 850~{\rm Myr}$, and $t\sim 1000~{\rm Myr}$, respectively.
See paper I in details.

In the $H_{\rm TT,5pc}$ model, the particle numbers of SPHs, pre-existing and old stars, and dark matter at $t=0$ are $442958$, $2051314$, and $27720000$, respectively.
The mass and gravitational softening length of SPH particles are $7.5\times 10^{3}~M_{\odot}$ and $5~\rm{pc}$, respectively.
In the simulation, multi-phase nature of ISM are realized by taking a wide temperature range of radiative cooling ($10~\rm{K}<T<10^{8}~\rm{K}$) and energy feedback from Type II supernovae (SNe) into account.
Metal contamination of gas by Type II SNe is also taken into account \citep{1994A&A...281L..97S}.
Star formations take place from cold ($T<100 ~\rm{K}$) and dense ($n_{\rm{H}}>100~\rm{cm}^{-3}$) gas.
The mass of a star particle spawned from an SPH is one third of SPH mass.

In addition to model $H_{\rm TT,5pc}$, the lower resolution models $H_{\rm TT}$ and $L_{\rm TT}$ are also analyzed.
These models have different mass and spatial resolutions from model $H_{\rm TT,5pc}$, but same collision parameter. 
In $H_{\rm TT}$ model, the gravitational softening length is $20~{\rm pc}$ but same SPH mass as $H_{\rm TT,5pc}$.
In $L_{\rm TT}$ model, SPH mass is $3\times 10^{4}~M_{\odot}$,
and gravitational softening length is $20~\rm{pc}$.

\subsection{Identification of young massive clusters}

We analyze simulation data at $t\sim 1350~{\rm Myr}$.
At that time, about $300~{\rm Myr}$ passes after the galaxy merger is completed.
We regard lower limit of detectable mass of the cluster as $2\times 10^5~{M_{\odot}}$ in $H_{\rm TT,5pc}$ and $H_{\rm TT}$ and as $8\times 10^5~{M_{\odot}}$ in $L_{\rm TT}$,
since clusters can be expressed by $100$ new star particles.

In order to detect YMCs, we firstly compute the gravitational potential energy of old and new stars.
After the calculation, we perform gravitational bound check for old and new star particles around the particle with locally minimum potential.
If particles are gravitationally bound and the total mass of the bound particles exceed the mass limit, we identify the system as a YMC.

\section{Results} \label{sec:results}

\subsection{Merger remnant}

\begin{figure}
 \begin{center}
  \includegraphics[width=80mm]{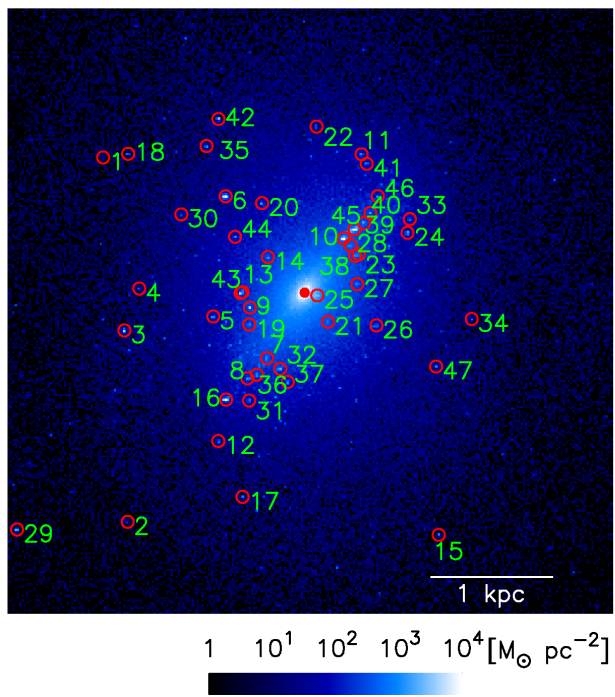}
 \end{center}
 \caption{Snapshot of both new stars and old stars in a merger remnant.
The size of the panel is $5~{\rm kpc}$.
The colors represent surface density of both new stars and old stars.
A filled and open circles denote the galactic center and detected YMCs.
The numbers around the YMCs show ID of clusters.
Although some faint dots can be observable in the other points, these objects cannot be identified as YMCs since they are gravitationally unbound or cluster mass does not reach the detection limit mass.
Two YMCs are not seen in this figure since these are much far from the galactic center.}\label{snapshot_galaxy}
\end{figure}

Snapshot of a merger remnant at $t\sim 1350~{\rm Myr}$ in $H_{\rm TT,5pc}$ model is shown in figure~\ref{snapshot_galaxy}.
We detect 49 YMCs in the remnant.

\begin{figure}
 \begin{center}
  \includegraphics[width=60mm]{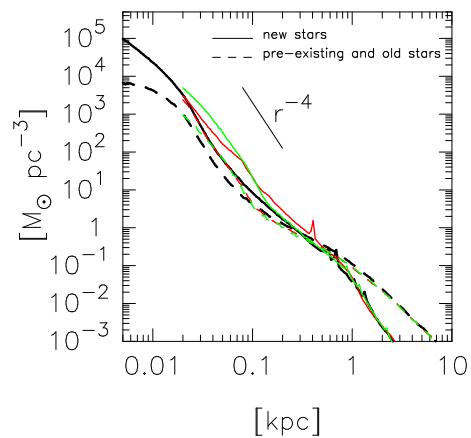}
 \end{center}
 \caption{Density profiles of young stars (solid lines) and pre-existing and old stars (dashed lines) in the merger remnants.
 Black, red, and green lines show simulation models of $H_\mathrm{TT,5pc}$, $H_\mathrm{TT}$, and $L_\mathrm{TT}$, respectively.
 The red and green lines are not drawn within $20~{\rm pc}$ since gravitational softening length, namely spatial resolution, is $20~{\rm pc}$.}\label{profile}
\end{figure}

Figure~\ref{profile} shows the density profile of new stars and pre-existing and old stars in the merger remnant at $t\sim 1350~\rm{Myr}$.
The density profile of new stars is steep and declines as approximately $\rho \propto r^{-4}$ in outer region larger than several ten parsec.
The profile with $r^{-4}$ is produced by experience of strong gravitational disturbances \citep{1990PASJ...42..205M}.
In the case of a galaxy merger, the intense disturbances are caused by merging of galactic cores and sink of hypermassive star clusters through dynamical friction.
Within $500~\rm{pc}$ from the galactic center, new stars are dominant.
The central density is much higher than that of both pre-existing and old stars by one order of magnitude.
Such structures are compatible with extremely compact structures of newly formed stars observed in ultraluminous infrared galaxies (ULIRGs) \citep{2000AJ....119..509S,2014MNRAS.440L..31M}.

The merger remnant has the total stellar mass of $1.2\times 10^{10}~M_{\odot}$ at that time.
$V$-band absolute magnitude $M_\mathrm{V}$ of the remnant is estimated by utilizing data of star particles and population synthesis code $PEGASE$ \citep{1999astro.ph.12179F}.
Then, $M_\mathrm{V} = -19.8$ is obtained, so that the specific frequency defined as $S_\mathrm{N}=N_\mathrm{cl} \times 10^{(M_\mathrm{V}+15)}$ \citep{1981AJ.....86.1627H} is $0.61$.
Here, $N_\mathrm{cl}$ is number of YMCs and $N_\mathrm{cl}=49$, and an effect of dust absorption is not taken into account.
This value seems to be lower than that observed in dwarf ellipticals and is comparable to that of late-type galaxies \citep{1998ApJ...508L.133M}.
This is because only $300~{\rm Myr}$ passes after galaxy-galaxy merging is completed and star particles are young so that the remnant is still bright.
If star particles evolve further and the remnant becomes dark, the value is close to observed one.
For example, when further $1000~{\rm Myr}$ passes, $M_{\rm V}$ and $S_\mathrm{N}$ becomes $-18.9$ and $1.4$, respectively, assuming passive evolution.

\subsection{Distribution of young massive clusters}

\begin{figure}
 \begin{center}
  \includegraphics[width=80mm]{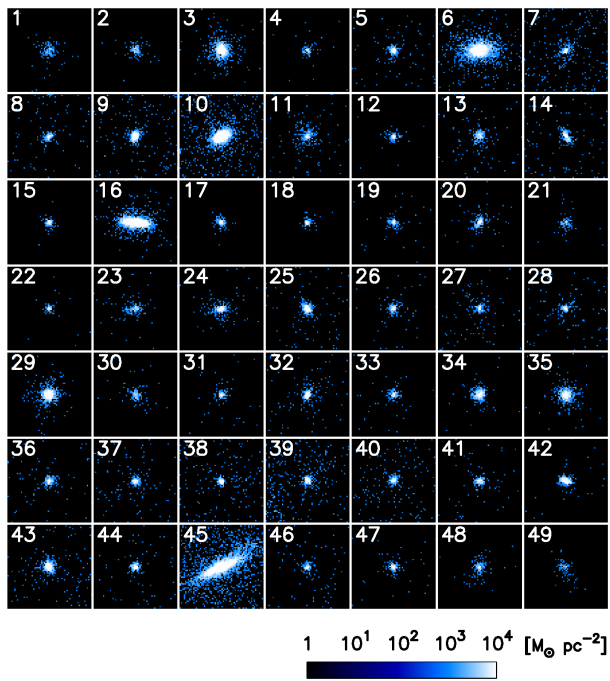}
 \end{center}
 \caption{Snapshots of detected YMCs.
 The size of each panel is $100~{\rm pc}$.
 The number displayed at upper left in each panel shows ID.}\label{snapshot_sc}
\end{figure}

\begin{figure}
 \begin{center}
  \includegraphics[width=80mm]{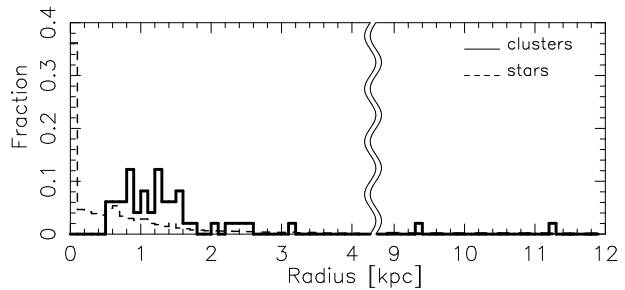}
 \end{center}
 \caption{Distribution of star clusters and new stars.
 The horizontal and vertical axes show radius from the center of the merger remnant and number fraction of them.
 The graph has a bin width of $100~{\rm pc}$.}\label{radius-fraction}
\end{figure}

All detected YMCs are shown in figure~\ref{snapshot_sc}.
The physical quantities of the clusters are listed in Tab.~\ref{tab_sc}.
Figure~\ref{radius-fraction} shows the fraction of YMCs and new stars in number as a function of radius from the galactic center.
Most of YMCs are located at the region of a few ${\rm kpc}$ from the galactic center
whereas new stars are extremely concentrated in the central region and nearly $40~\%$ of them are distributed within $100~{\rm pc}$.
In $H_{\rm TT}$ and $L_{\rm TT}$ cases, distribution of YMCs and new stars is similar to that of $H_{\rm TT,5pc}$ case as shown in figure~\ref{profile} and table~\ref{tab_sc2}.
The difference of distribution between YMCs and stars in the galactic central region is consistent with observations \citep{2006A&A...458...53S} and can be explained as follows.
YMCs form mainly from widespread gas filaments generated at encounter of two galaxies or spatially extended turbulent gas generated by its inflow toward galactic central region at final coalescence phase of galactic cores.
On the other hand, a large number of new stars are supplied by sinking of hypermassive star clusters in addition to star formations from such compressed gas (paper I).
The formation epoch of YMCs in our simulations is different from \citet{2015MNRAS.446.2038R} in which cluster formations are mainly take place at the first encounter of galaxies.
This would be due to difference of collision parameters.
In \citet{2015MNRAS.446.2038R}, Antennae-like collision parameter is adopted, which reproduces compression of gas and formation of star clusters after the first encounter of galaxies.

Two of clusters, of which ID is 48 and 49, appear at $\sim 10~{\rm kpc}$ from the center of the merger remnant.
These clusters are gravitationally bound by the remnant.
Their formations take place in tidal tails apart from the center, which is different from formations of other clusters.
These are classified as tidal dwarfs.
In our simulations, there is no ejected YMC from the galaxy during galaxy-galaxy interaction as suggested by \citet{2010ApJ...712L.184E}.
The difference of the formation site produces two component of YMC distribution.
In $H_{\rm TT}$ and $L_{\rm TT}$ cases, formation of a tidal dwarf does not occur.

\begin{figure}
 \begin{center}
  \includegraphics[width=50mm]{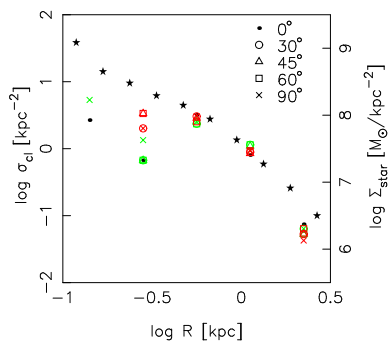}
 \end{center}
 \caption{Surface number density $\sigma _{\rm cl}$ of YMCs and surface new stellar mass density $\Sigma _{\rm star}$ as a function of radius from the galactic center.
 The bin width of $\sigma _{\rm cl}$ and $\Sigma _{\rm star}$ is 0.3, 0.15, respectively.
 In order to obtain projected snapshots from various viewpoints, we rotate figure~\ref{snapshot_galaxy} with respect to its horizontal or vertical axis.
 The red and green plots show the horizontal and vertical cases, respectively.
 The dots, circles, triangles, squares, and crosses denote $\sigma _{\rm cl}$ in $0^{\circ}$, $30^{\circ}$, $45^{\circ}$, $60^{\circ}$, and $90^{\circ}$ cases, respectively.
 The black star symbols denote $\Sigma _{\rm star}$.
 }\label{surface}
\end{figure}

In order to compare our data with observations, we show the surface number density of YMCs as a function of radius from the galactic center in figure~\ref{surface}.
Here, the surface density is calculated by using projected snapshots which are observed from various viewpoints.
The figure shows that the distribution of YMCs do not strongly depend on viewing angle.
Whereas the stellar distribution traces distribution of YMCs in the galactic outer region, the deficit of YMCs appears in the galactic inner region.
This result is in agreement with observations of isolated elliptical galaxies \citep{2006A&A...458...53S,2015A&A...577A..59S}.

\subsection{Mass function of young massive clusters}

\begin{figure}
 \begin{center}
  \includegraphics[width=70mm]{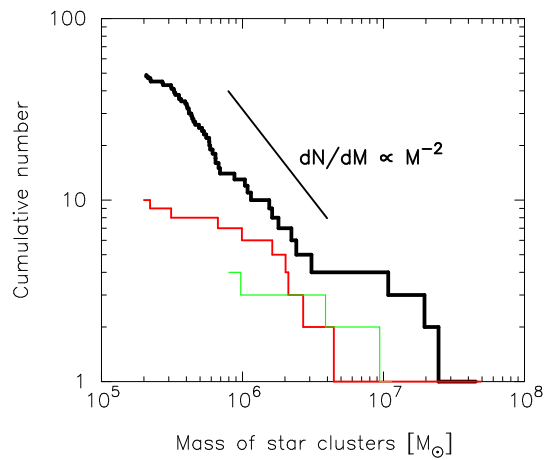}
 \end{center}
 \caption{Cumulative mass function. The horizontal and vertical axes are cumulative number and mass of YMCs, respectively.
 The black, red, green lines show $H_\mathrm{TT,5pc}$, $H_\mathrm{TT}$, $L_\mathrm{TT}$, respectively.}\label{mass_fanction}
\end{figure}

Cumulative mass function of YMCs is shown in figure~\ref{mass_fanction}.
In $H_\mathrm{TT,5pc}$ model, the masses of clusters range from $2.08\times 10^5~M_{\odot}$ to $4.51 \times 10^7~M_{\odot}$.
There are no hypermassive star clusters with $\sim 10^8~M_{\odot}$ since they have already sunk to the galactic center through dynamical friction.
This graph indicates that mass function becomes a power low function with $dN/dM \propto M^{-2}$,
which is good agreement with observations \citep{1999AJ....118.1551W} and previous numerical simulations \citep{2010ASPC..423..185S}.

\begin{figure}
 \begin{center}
  \includegraphics[width=80mm]{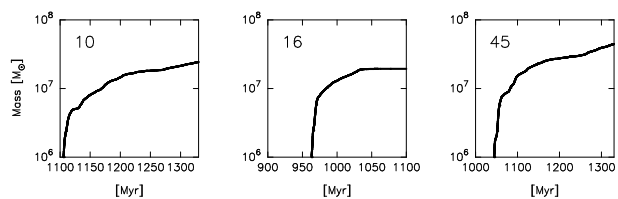}
 \end{center}
 \caption{Time evolution of stellar mass of massive YMCs, of which IDs are 10, 16, and 45 respectively.}\label{mass_evolution}
\end{figure}

A deviation from the line of $dN/dM \propto M^{-2}$, however, appears at $\sim 10^{7}~M_{\odot}$, in other words, there is an excess of massive clusters.
The stellar mass evolutions of massive YMCs, of which IDs are 10, 16, and 45, are shown in figure~\ref{mass_evolution}.
The figure shows that their initial stellar masses are less than $\sim 10^{7}~M_{\odot}$.
After formations of YMCs, these 3 YMCs continue to pass filamentary dense gas regions which exist within $\sim 1~{\rm kpc}$ from the galactic center.
The YMCs obtain such dense gas continuously, and formations of new stars take place within YMCs.
As a result, stellar mass growth of the YMCs occurs and the excess of massive YMCs emerges.
Since the cluster with ID~16 escapes from the dense gas region, mass growth is quenched at $t\sim 1030~{\rm Myr}$.
These objects seem to be spatially extended (see figure \ref{snapshot_sc}) than observed young clusters which have half light radius with a few pc \citep{2008A&A...489.1091M}.
Such massive clusters would sink into the galactic center through dynamical friction less than $1~{\rm Gyr}$ according to Chandrasekahr's formula

\begin{equation}
t_{\rm dyn} = \frac{0.95~{\rm Gyr}}{\ln \Lambda} \Big( \frac{r}{0.5~{\rm kpc}} \Big)^2 \frac{\sigma}{200~{\rm km~s^{-1}}} \frac{2\times 10^7~M_{\odot}}{M}
\end{equation}

\noindent \citep{1987gady.book.....B}. Then, the excess disappears and spatially extended clusters would not been observed.

In $H_\mathrm{TT}$ model, mass function in the less massive region than $10^{6}~M_{\odot}$ is deviated from $dN/dM \propto M^{-2}$.
This is because large gravitational softening length prevents from formations of clusters less than $10^{6}~M_{\odot}$.

\subsection{Age of young massive clusters}

In table~\ref{tab_sc}, $t_{{\rm XX}\%}$ indicates the time when ${\rm XX}\%$ of stars in each cluster are formed.
According to $t_{25\%}$, $t_{\rm 50\%}$, and $t_{\rm 75\%}$, the clusters are able to be classified into the following three types:
(1) clusters consisting of only new stars with single age, (2) clusters consisting of only new stars with bimodal distribution in age, (3) massive clusters consisting of only new stars with various age, and (4) clusters constituting of both new and old stars.

\begin{figure}
 \begin{center}
  \includegraphics[width=85mm]{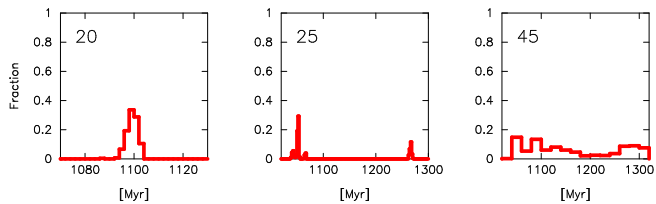}
 \end{center}
 \caption{Distribution of star formation time in each YMC formed after the galaxy encounter. The horizontal and vertical axes are star formation time and number fraction of stars to total stellar number, respectively.
 The number at upper-left denotes cluster ID.}\label{age1}
\end{figure}

Thirty eight YMCs consist of only new stars with single age.
All of these clusters form after galaxy-galaxy encounter,
and their formations are induced by the galaxy merger.
The typical population of stellar age within these clusters is shown in the left panel of figure~\ref{age1}
which shows number fraction of stars within a cluster as a function of star formation time.
Stars within these clusters have similar formation epoch and the width of age distribution is $10~{\rm Myr}$ which is timescale of feedback to eject gas.

\begin{figure}
 \begin{center}
  \includegraphics[width=85mm]{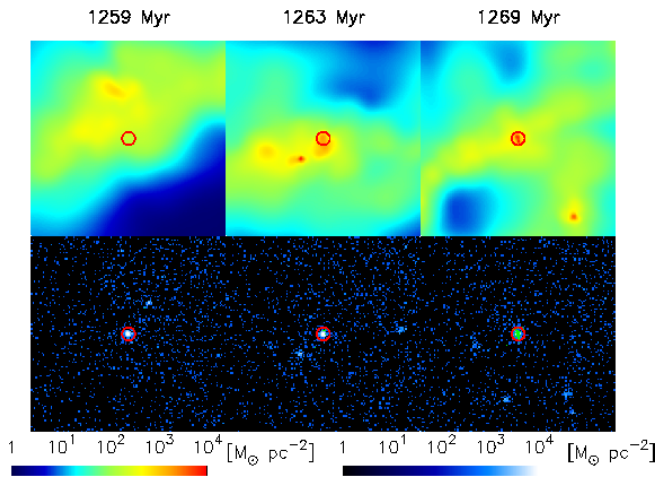}
 \end{center}
 \caption{Snapshots of the cluster with ID~25 from $t=1259~{\rm Myr}$ to $t=1269~{\rm Myr}$.
 Above and bottom panels show surface density of gas and new star, respectively.
 The red circle represents the position of the cluster.
 In bottom panels, green dots represent formed next generation stars.
 At $t=1259$, dense gas region appears above the cluster.
 After that, the cluster pass through this region.
 Then, the cluster captures gas at $t=1263~{\rm Myr}$ and star formation takes place within the cluster.}\label{snap25}
\end{figure}

One YMC, of which ID is $25$, has bimodal distribution in age as shown in the middle panel of figure~\ref{age1}.
The interval of age between two events is $200~{\rm Myr}$.
The process of producing such population is as follows.
The cluster forms at $t=1050~{\rm Myr}$.
After the formation, the cluster wanders few gas region.
When the cluster passes the dense gas region temporarily after $t\sim 1250~{\rm Myr}$, it captures gas as shown in the left and middle panels of figure~\ref{snap25}.
After that, the density of gas within the cluster becomes high and gas cools because of radiative cooling.
As a result, the next generation stars form as shown in the right panel of figure~\ref{snap25}.

Three YMCs, of which IDs are $10$, $16$, and $45$, exceed $10^{7}~M_{\odot}$.
These clusters consist of stars with various ages.
The typical distribution of the stellar formation time is shown in the right panel of figure~\ref{age1}.
The distribution is rather continuous than discrete unlike ID $25$.
This is because these YMCs continue to pass dense gas region and obtain gas continuously.
The continuous gas accretion results in star formations within the clusters at various time.

\begin{figure}
 \begin{center}
  \includegraphics[width=85mm]{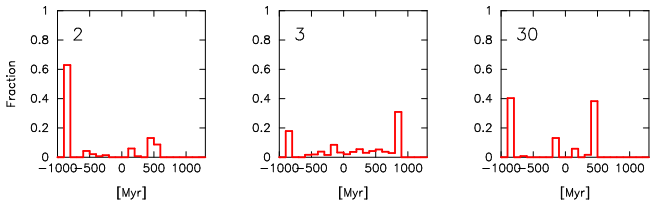}
 \end{center}
 \caption{Same as figure~\ref{age1} but clusters formed before the galaxy encounter.}\label{age2}
\end{figure}

The other 7 YMCs consist of old and new stars although clusters, of which IDs are $1$ and $34$, have few new stars.
All of these 7 clusters form in an initial unstable gas disk at $t\sim -800~{\rm Myr}$ during a simulation of an isolated gas disk.
Figure~\ref{age2} shows distribution of formation time of stars within these clusters.
Although these clusters form in an initial isolated unstable gas disk at $t\sim -800~{\rm Myr}$, they include stars with various formation time.
This is because clusters capture gas when they pass through dense gas region and star formations occur within them similarly to the cluster of ID 25.
These clusters are likely to contain stars formed at the encounter phase, namely around $400~{\rm Myr}$ or $800~{\rm Myr}$.
The reason is that the encounter produces turbulent and dense gas and increases the probability that a cluster passes through the dense gas region.

In $H_{\rm TT}$ and $L_{\rm TT}$ cases, $t_{25\%}$, $t_{\rm 50\%}$, and $t_{\rm 75\%}$ are shown in table~\ref{tab_sc2}.
All YMCs form after the second encounter of galaxies.
Whereas most YMCs have single stellar population, ID~$50$, $61$, and $62$ has multiple stellar populations.
ID~$50$ and $62$ consist of stars with various ages similarly to ID~$10$, $16$, and $45$, and ID~$61$ has bimodal stellar population in age similarly to ID~$25$.
The multiple population is formed by capturing dense gas and forming the second generation stars similarly to the $H_{\rm TT,5pc}$ case.

\begin{figure}
 \begin{center}
  \includegraphics[width=40mm]{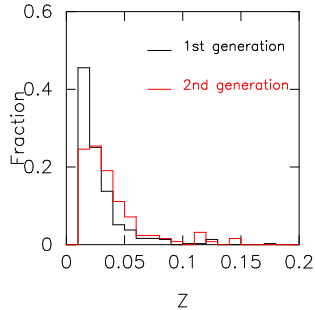}
 \end{center}
 \caption{The number fraction of stars as a function of metallicity $Z$ in the cluster with ID~25.
 The black and red lines show the first and the second generation stars, respectively.
 The bin width is $0.01$.}\label{metallicity}
\end{figure}

Recent observations have revealed that Galactic globular clusters generally have multiple stellar populations \citep{2017arXiv171201286B}.
Such globular clusters can be divided into two types in accordance with abundance patterns.
The first and second types contain Fe and light element (e.g., He, C, N, O, Na, and Al) variations, respectively.
Since we do not have abundance information for stars in our simulation, we conjecture types of YMCs with bimodal distribution in age.
These YMCs should belong to the first type in the following reason.
These YMCs except ID~25 contains stellar populations formed before and after the galaxy-galaxy merger.
These stellar populations clearly have different Fe abundance.
The YMC with ID~25 has stellar populations formed at $t\sim 1050~{\rm Myr}$ and $t\sim 1250~{\rm Myr}$.
The metallicity, $Z$, of the first and second generation stars in the YMC with ID~25 is shown in figure~\ref{metallicity}.
The figure shows that metallicity of the second generation stars increases by $\sim 0.01$ compared to that of first generation stars.
Although the enrichment seems to be mild, it becomes more remarkable in high-z galaxies with low metallicity.
The difference of metallicity between the first and the second generation stars indicates that the second generation stars form from gas contaminated by TypeII SNe.
Thus, these clusters correspond to $\omega$ Cen with different Fe abundance.
On the other hand, the second type is expected to be formed from asymptotic giant branch (AGB) ejecta of the first generation stars \citep{2008MNRAS.391..354R,2017MNRAS.471.2242B,2018arXiv180702309B}.

\renewcommand{\arraystretch}{0.7} 
\begin{table*}
  \tbl{Property of YMCs in the $H_{\rm TT, 5pc}$ case.}{%
  \begin{tabular}{ccccccccc}
      \hline
	ID & $M$ [$M_{\odot}$]\footnotemark[$*$] & $r$ [${\rm kpc}$]\footnotemark[$\dag$]  & $t_{25\%}$ [${\rm Myr}$]\footnotemark[$\ddag$]  & $t_{50\%}$ [${\rm Myr}$]\footnotemark[$\S$] & $t_{75\%}$ [${\rm Myr}$] \footnotemark[$\|$]
	& $r_\mathrm{p}$\footnotemark[$\sharp$] & $r_\mathrm{a}$\footnotemark[$**$]  & $\epsilon$\footnotemark[$\dag\dag$] \\ 
      \hline
1 & $3.14\times 10^5$ & $2.01$ & $-779$ & $-778$ & $-703$ & $1.88$ & $2.03$ & $0.04$ \\
2 & $3.12\times 10^5$ & $2.40$ & $-845$ & $-843$ & $146$ & $1.58$ & $2.63$ & $0.25$ \\
3 & $3.09\times 10^6$ & $1.55$ & $-369$ & $298$ & $832$ & $0.65$ & $1.69$ & $0.45$ \\
4 & $2.23\times 10^5$ & $1.38$ & $742$ & $743$ & $744$ & $1.37$ & $1.82$ & $0.14$ \\
5 & $6.43\times 10^5$ & $0.81$ & $1061$ & $1062$ & $1063$ & $0.44$ & $0.85$ & $0.31$ \\
6 & $1.08\times 10^7$ & $1.08$ & $978$ & $984$ & $987$ & $0.62$ & $1.29$ & $0.35$ \\
7 & $4.41\times 10^5$ & $0.64$ & $1238$ & $1239$ & $1240$ & $0.55$ & $0.68$ & $0.11$ \\
8 & $4.02\times 10^5$ & $0.88$ & $1182$ & $1183$ & $1185$ & $0.59$ & $0.90$ & $0.21$ \\
9 & $2.22\times 10^6$ & $0.54$ & $1030$ & $1030$ & $1032$ & $0.53$ & $0.88$ & $0.25$ \\
10 & $2.46\times 10^7$ & $0.58$ & $1136$ & $1180$ & $1254$ & $0.58$ & $0.59$ & $0.01$ \\
11 & $5.53\times 10^5$ & $1.24$ & $1181$ & $1182$ & $1183$ & $1.05$ & $1.32$ & $0.11$ \\
12 & $3.68\times 10^5$ & $1.58$ & $1151$ & $1153$ & $1154$ & $0.98$ & $1.60$ & $0.24$ \\
13 & $4.34\times 10^5$ & $0.83$ & $967$ & $970$ & $971$ & $0.79$ & $1.38$ & $0.28$ \\
14 & $5.90\times 10^5$ & $0.71$ & $1068$ & $1069$ & $1070$ & $0.60$ & $0.74$ & $0.10$ \\
15 & $4.17\times 10^5$ & $2.39$ & $1033$ & $1034$ & $1035$ & $0.57$ & $2.40$ & $0.62$ \\
16 & $1.95\times 10^7$ & $1.41$ & $969$ & $984$ & $1011$ & $0.25$ & $1.70$ & $0.75$ \\
17 & $3.93\times 10^5$ & $2.51$ & $952$ & $953$ & $954$ & $1.02$ & $2.54$ & $0.43$ \\
18 & $3.28\times 10^5$ & $2.29$ & $705$ & $706$ & $707$ & $0.95$ & $2.72$ & $0.48$ \\
19 & $3.26\times 10^5$ & $1.21$ & $1267$ & $1268$ & $1269$ & $0.49$ & $1.28$ & $0.45$ \\
20 & $6.14\times 10^5$ & $1.41$ & $1098$ & $1099$ & $1101$ & $0.83$ & $1.46$ & $0.27$ \\
21 & $2.22\times 10^5$ & $1.36$ & $1093$ & $1102$ & $1105$ & $0.65$ & $1.48$ & $0.39$ \\
22 & $2.14\times 10^5$ & $1.73$ & $1058$ & $1059$ & $1060$ & $1.21$ & $1.86$ & $0.21$ \\
23 & $3.55\times 10^5$ & $1.24$ & $1278$ & $1279$ & $1280$ & $0.98$ & $1.24$ & $0.12$ \\
24 & $1.04\times 10^6$ & $1.57$ & $1100$ & $1102$ & $1104$ & $0.96$ & $1.58$ & $0.25$ \\
25 & $1.15\times 10^6$ & $0.88$ & $1051$ & $1053$ & $1259$ & $0.83$ & $0.89$ & $0.04$ \\
26 & $4.62\times 10^5$ & $1.17$ & $1166$ & $1168$ & $1169$ & $0.84$ & $1.23$ & $0.19$ \\
27 & $3.54\times 10^5$ & $1.05$ & $1266$ & $1269$ & $1270$ & $0.87$ & $1.16$ & $0.15$ \\
28 & $5.17\times 10^5$ & $1.04$ & $1318$ & $1319$ & $1321$ & $0.89$ & $1.12$ & $0.11$ \\
29 & $2.41\times 10^6$ & $3.17$ & $-558$ & $14$ & $487$ & $0.61$ & $3.39$ & $0.70$ \\
30 & $2.71\times 10^5$ & $1.32$ & $-839$ & $-111$ & $416$ & $1.26$ & $3.78$ & $0.50$ \\
31 & $2.74\times 10^5$ & $1.16$ & $1154$ & $1155$ & $1157$ & $0.82$ & $1.23$ & $0.20$ \\
32 & $8.76\times 10^5$ & $0.91$ & $1147$ & $1149$ & $1150$ & $0.91$ & $1.04$ & $0.07$ \\
33 & $4.16\times 10^5$ & $1.24$ & $1115$ & $1116$ & $1117$ & $0.73$ & $1.24$ & $0.26$ \\
34 & $1.09\times 10^6$ & $1.50$ & $-872$ & $-870$ & $-869$ & $0.25$ & $1.67$ & $0.74$ \\
35 & $1.55\times 10^6$ & $1.50$ & $-871$ & $-869$ & $472$ & $0.92$ & $2.88$ & $0.52$ \\
36 & $5.33\times 10^5$ & $0.88$ & $1182$ & $1184$ & $1185$ & $0.62$ & $0.91$ & $0.19$ \\
37 & $5.81\times 10^5$ & $0.78$ & $1062$ & $1063$ & $1064$ & $0.59$ & $0.82$ & $0.16$ \\
38 & $6.77\times 10^5$ & $0.62$ & $1197$ & $1198$ & $1199$ & $0.31$ & $0.71$ & $0.40$ \\
39 & $6.44\times 10^5$ & $0.80$ & $1231$ & $1232$ & $1234$ & $0.58$ & $0.88$ & $0.21$\\
40 & $5.82\times 10^5$ & $0.90$ & $1060$ & $1061$ & $1061$ & $0.67$ & $0.95$ & $0.17$ \\
41 & $4.49\times 10^5$ & $1.21$ & $1169$ & $1171$ & $1172$ & $1.13$ & $1.29$ & $0.06$ \\
42 & $1.62\times 10^6$ & $1.62$ & $950$ & $950$ & $951$ & $0.32$ & $3.47$ & $0.83$ \\
43 & $1.80\times 10^6$ & $0.53$ & $1017$ & $1019$ & $1019$ & $0.36$ & $0.91$ & $0.44$ \\
44 & $6.93\times 10^5$ & $0.74$ & $1022$ & $1023$ & $1023$ & $0.26$ & $1.59$ & $0.72$ \\
45 & $4.51\times 10^7$ & $0.69$ & $1087$ & $1145$ & $1269$ & $0.69$ & $0.78$ & $0.06$ \\
46 & $3.34\times 10^5$ & $1.02$ & $1176$ & $1177$ & $1178$ & $0.94$ & $1.39$ & $0.19$ \\
47 & $4.91\times 10^5$ & $1.25$ & $875$ & $876$ & $876$ & $0.62$ & $1.42$ & $0.39$ \\
48 & $4.07\times 10^5$ & $11.30$ & $817$ & $819$ & $820$ & $11.04$ & $18.67$ & $0.26$ \\
49 & $2.08\times 10^5$ & $9.31$ & $762$ & $765$ & $766$ & $9.07$ & $17.63$ & $0.32$ \\
      \hline
    \end{tabular}}\label{tab_sc}
\begin{tabnote}
\footnotemark[$*$] Mass of a YMC.  \\ 
\footnotemark[$\dag$] Distance of a YMC from the center of the merger remnant. \\
\footnotemark[$\ddag$] Star formation time of a star located at $25\%$ when stars within a YMC are sorted from young to old. \\ 
\footnotemark[$\S$] Star formation time of a star located at $50\%$ when stars within a YMC are sorted from young to old.\\ 
\footnotemark[$\|$] Star formation time of a star located at $75\%$ when stars within a YMC are sorted from young to old.\\
\footnotemark[$\sharp$]  Pericenter distance of a YMC. \\  
\footnotemark[$**$]   Apocenter distance of a YMC.\\ 
\footnotemark[$\dag\dag$]  Eccentricity of an orbit. \\ 
\end{tabnote}
\end{table*}

\begin{table*}
  \tbl{Property of YMCs in $H_{\rm TT}$ and $L_{\rm TT}$ cases. YMCs from ID~50 to ID~59 and from ID~60 to ID~63 form in $H_{\rm TT}$ and $L_{\rm TT}$ cases, respectively.}{%
  \begin{tabular}{ccccccccc}
      \hline
	ID & $M$ [$M_{\odot}$]\footnotemark[$*$] & $r$ [${\rm kpc}$]\footnotemark[$\dag$]  & $t_{25\%}$ [${\rm Myr}$]\footnotemark[$\ddag$]  & $t_{50\%}$ [${\rm Myr}$]\footnotemark[$\S$] & $t_{75\%}$ [${\rm Myr}$] \footnotemark[$\|$]
	& $r_\mathrm{p}$\footnotemark[$\sharp$] & $r_\mathrm{a}$\footnotemark[$**$]  & $\epsilon$\footnotemark[$\dag\dag$] \\ 
      \hline
50 & $4.90\times 10^7$ & $0.41$ & $898$ & $917$ & $969$ & $0.38$ & $1.28$ & $0.54$ \\
51 & $2.21\times 10^5$ & $1.15$ & $1173$ & $1175$ & $1177$ & $1.09$ & $1.22$ & $0.06$ \\
52 & $2.02\times 10^6$ & $1.06$ & $1164$ & $1166$ & $1168$ & $1.05$ & $1.10$ & $0.02$ \\
53 & $6.69\times 10^5$ & $0.73$ & $1195$ & $1197$ & $1198$ & $0.69$ & $0.73$ & $0.03$ \\
54 & $1.63\times 10^6$ & $2.97$ & $897$ & $901$ & $903$ & $0.08$ & $3.03$ & $0.95$ \\
55 & $9.90\times 10^5$ & $1.37$ & $917$ & $918$ & $919$ & $0.77$ & $1.43$ & $0.30$ \\
56 & $4.45\times 10^6$ & $1.16$ & $889$ & $897$ & $900$ & $0.75$ & $1.46$ & $0.32$ \\
57 & $2.69\times 10^6$ & $0.84$ & $1000$ & $1001$ & $1002$ & $0.34$ & $0.87$ & $0.43$ \\
58 & $3.12\times 10^5$ & $1.19$ & $888$ & $895$ & $899$ & $0.75$ & $1.44$ & $0.31$ \\
59 & $2.11\times 10^6$ & $0.95$ & $1025$ & $1026$ & $1027$ & $0.94$ & $1.08$ & $0.07$ \\
60 & $9.40\times 10^6$ & $1.21$ & $1011$ & $1015$ & $1020$ & $0.82$ & $1.21$ & $0.19$ \\
61 & $1.14\times 10^7$ & $0.91$ & $1200$ & $1203$ & $1314$ & $0.75$ & $0.91$ & $0.10$ \\
62 & $3.88\times 10^6$ & $2.24$ & $883$ & $885$ & $913$ & $0.17$ & $3.44$ & $0.91$ \\
63 & $9.72\times 10^5$ & $0.90$ & $1214$ & $1215$ & $1217$ & $0.90$ & $1.13$ & $0.11$ \\
      \hline
    \end{tabular}}\label{tab_sc2}
\begin{tabnote}
\footnotemark[$*$] Mass of a YMC.  \\ 
\footnotemark[$\dag$] Distance of a YMC from the center of the merger remnant. \\
\footnotemark[$\ddag$] Star formation time of a star located at $25\%$ when stars within a YMC are sorted from young to old. \\ 
\footnotemark[$\S$] Star formation time of a star located at $50\%$ when stars within a YMC are sorted from young to old.\\ 
\footnotemark[$\|$] Star formation time of a star located at $75\%$ when stars within a YMC are sorted from young to old.\\
\footnotemark[$\sharp$]  Pericenter distance of a YMC. \\  
\footnotemark[$**$]   Apocenter distance of a YMC.\\ 
\footnotemark[$\dag\dag$]  Eccentricity of an orbit. \\ 
\end{tabnote}
\end{table*}

\subsection{Motion of young massive clusters}

In order to investigate orbits of YMCs in the merger remnant, we simulate motion of YMCs by the second order leapfrog integration scheme.
The gravitational potential of the remnant is produced by using distribution of SPH, star, and dark matter particles at $t\sim 1350~{\rm Myr}$.
Here, we assume that the gravitational potential is steady and spherically symmetry.
These assumptions are reasonable to calculate orbits approximately since asymmetry of the remnant is not strong and the remnant is quasi-stable.
Note that evaporation of YMCs and dynamical friction are not taken account.

The pericenter distance $r_\mathrm{p}$, apocenter distance $r_\mathrm{a}$, and eccentricity defined by $\epsilon = (r_\mathrm{a} - r_\mathrm{p})/(r_\mathrm{a} + r_\mathrm{p})$ are summarized
in table~\ref{tab_sc} for the high resolution case and in table~\ref{tab_sc2} for the low resolution cases.
In both cases, orbits of YMCs in the galactic inner region are various, but the number fraction of circular orbits is large.
This is because YMCs formed in a major merger do not lose the orbital angular momentum sufficiently.
This trend seems to be different from observations of Milky Way \citep{2018arXiv180409381G} or cosmological simulations of a disk galaxy \citep{2006ApJ...640...22S}
which is different morphology from a galaxy-galaxy merger remnant and do not experience a recent major merger.
Such observations and simulations show that a large fraction of YMCs have radial orbits.
This might indicate that formation process of clusters is different between a merging galaxy and a disk galaxy.
While YMCs form from filamentary gas or turbulent gas generated by a galaxy-galaxy merger,  YMCs form by in situ formation, minor mergers, and satellite accretion \citep{2017MNRAS.465.3622R}.

Eccentricities of orbits of the galactic outer YMCs, of which ID are 48 and 49, are not high.
This is because they form in rotating tidal tails and hence have originally sufficient orbital angular momentum.
These clusters migrate from $\sim 10~{\rm kpc}$ to $\sim 20~{\rm kpc}$.
Since their pericenter and apocenter distance are $\sim 10~{\rm kpc}$ and $\sim 20~{\rm kpc}$, respectively, these objects are observed as isolated globular clusters.
The isolated globular clusters are observed in some galaxies, for examples M31 \citep{2010MNRAS.401..533M} and M81 \citep{2012ApJ...751L..19J}, although it is difficult to compare simply with spiral galaxies.

\section{Summary}

We investigate properties of YMCs in a merger remnant by analyzing simulation data in paper I.
Our findings are as follows.
\begin{itemize}
\item~YMCs are formed in filamentary and turbulent gas and tidal tails generated by a galaxy merger.
The former and the latter are distributed at a few $\rm kpc$ and at $\sim 10~{\rm kpc}$ from the center of the merger remnant, respectively.
The YMCs are much less concentrated than galaxy stars.
\item~The mass function of YMCs becomes $dN/dM \propto M^{-2}$, but the excess appears around $10^{7}~M_{\odot}$ after merging of galactic cores is completed.
The excess would disappear less than $1~{\rm Gyr}$ due to dynamical friction.
\item~Most of YMCs formed during a galaxy merger consist of single stellar population in age.
On the other hand, the rest have multiple distribution in age.
The multiple population is formed by capturing dense gas and forming new stars within a YMC when a YMC pass dense gas region.
\item~Orbits of YMCs in the inner galactic region are various, but large fraction of candidates would rather have circular orbits.
Eccentricities of YMCs in the inner galactic region are not high.
\end{itemize}

\bigskip

We thank Dr. Florent Renaud for giving us useful comments.
Numerical computations were carried out on Cray XT4 at Center for Computational Astrophysics (CfCA), National Astronomical Observatory of Japan,
and numerical analyses were carried out on computers at CfCA, National Astronomical Observatory of Japan.
This research has been supported in part by the MEXT programme for the Development and Improvement for the Next Generation Ultra High-Speed Computer System
under its Subsidies for Operating the Specific Advanced Large Research Facilities,
and by Grants-in-Aid for Scientific Research (16K17656, 17H06360) from the Japan Society for the Promotion of Science.


\end{document}